\documentclass[fleqn,10pt]{wlscirep}

\usepackage[normalem]{ulem}
\usepackage[utf8]{inputenc}
\usepackage[T1]{fontenc}
\usepackage{physics}
\usepackage{graphicx}
\usepackage{hyperref}
\usepackage[all]{hypcap}
\usepackage[caption=false]{subfig}

\title{Spectral compression and entanglement reduction in the cascaded biphoton state with cavities}

\author[1]{Y.-E Wong}
\affil[1]{Department of Mechanical Engineering, National Taiwan University, Taipei 106319, Taiwan}

\author[2]{N.-Y. Tsai}
\author[2]{W. S. Hiew}
\affil[2]{Department of Physics, National Taiwan University, Taipei 10617, Taiwan}

\author[3,*]{H. H. Jen}
\affil[3]{Institute of Atomic and Molecular Sciences, Academia Sinica, Taipei 10617, Taiwan}
\affil[*]{sappyjen@gmail.com}

\renewcommand{\r}{\mathbf{r}}

\renewcommand{\k}{\mathbf{k}}

\def\bea{\begin{eqnarray}}
\def\eea{\end{eqnarray}}
\def\ba{\begin{array}}
\def\ea{\end{array}}
\def\bdm{\begin{displaymath}}
\def\edm{\end{displaymath}}

\begin{abstract} 
The cascaded biphoton state generated from a cold atomic ensemble presents one of the strongly correlated resources that can preserve and relay quantum information. Under the four-wave mixing condition, the emitted signal and idler photons from the upper and lower excited states become highly correlated in their traveling directions and entangled in continuous frequency spaces. In this system, we theoretically study the spectral compression of the biphoton source using an external cavity and show the reduction in its frequency entanglement entropy. This indicates, respectively, an improved light absorption efficiency for the idler photon as well as an almost pure biphoton source which is useful in optical quantum networks. We further investigate the limit of the spectral compression that can be achieved by using multiple cavities. Our results show the capability and potential of the biphoton source with external cavities, where the performance of atom-based quantum memory can be enhanced and the entanglement property can be manipulated by tailoring the spectral compression.      
\end{abstract}
\begin{document}
\flushbottom
\maketitle
\section*{Introduction}

Long-distance quantum communication \cite{Duan2001} has progressed into a satellite-based realm \cite{Yin2017, Liao2017, Liao2018, Chen2021}, where a global scale of quantum network \cite{Kimble2008} can be achieved by integrating metropolitan-area networks via fiber links with the ground-satellite links \cite{Chen2021}. Conventional approach to transmit quantum information to far distance utilizes quantum repeater protocols \cite{Briegel1998, Dur1999}, which relay the entanglement and fulfill the distribution of it by inserting multiple entangled pairs in between with conditional local measurements. These protocols require efficient light-matter interfaces \cite{Hammerer2010, Chen2013, Yang2016, Hsiao2018}, which can be further improved by using telecommunication wavelengths \cite{Chaneliere2006, Radnaev2010, Jen2010, Albrecht2014} and multiplexing quantum memories in space \cite{Collins2007, Lan2009}, time \cite{Simon2007} or frequencies \cite{Bernhard2013, Lukens2014, Jen2016a, Jen2016b, Lukens2017, Kues2017}. These respectively allow a better transmission efficiency in fibers and a higher capacity in entanglement distribution \cite{Afzelius2009, Dai2012, Zheng2015}, which together lay the foundation toward a robust long-distance quantum communication. 

The photons in telecom bandwidth (signal) of $1.3$--$1.5$ $\mu$m can be generated from the upper excited state of the cascade transitions in rubidium and cesium atoms \cite{Chaneliere2006}, which are highly correlated to the subsequently emitted infrared photons (idler) under the four-wave mixing conditions. This correlated biphoton source further presents a superradiant signal-idler photon-photon correlation \cite{Jen2012, Jen2015} owing to the collective dipole-dipole interactions in a dense medium \cite{Dicke1954,  Lehmberg1970, Gross1982}. This makes a mismatch of the spectral width of the generated idler photons to the one of the absorbing quantum interface \cite{Hammerer2010}, which reduces the performance of quantum information transfer at the interface. This issue can be resolved either by using narrowband single photons from electromagnetically induced transparency in the atoms \cite{Eisaman2005} or enabling the spectral compression with a resonant cavity \cite{Seidler2020}.  

Here we theoretically investigate the spectral properties of the cascaded biphoton source under the external cavities. The spectral compression along with the reduction of the continuous entanglement entropy can be accomplished by dispersing and modulating the idler photons with a near-resonant and lossless cavity. This presents an active manipulation of the spectral width in the biphoton source, which can be tailored to match the resonance width of the atomic transition. In doing so, the reduction of the bipartite entanglement entropy in frequency spaces indicates an almost pure single photon sources, which is useful to implement linear optical quantum networks \cite{Dusanowski2019}.  

\section*{Results}

\subsection*{Modulated spectral function of biphoton source}

\begin{figure}[t]
\centering
\includegraphics[width=12cm,height=8.5cm]{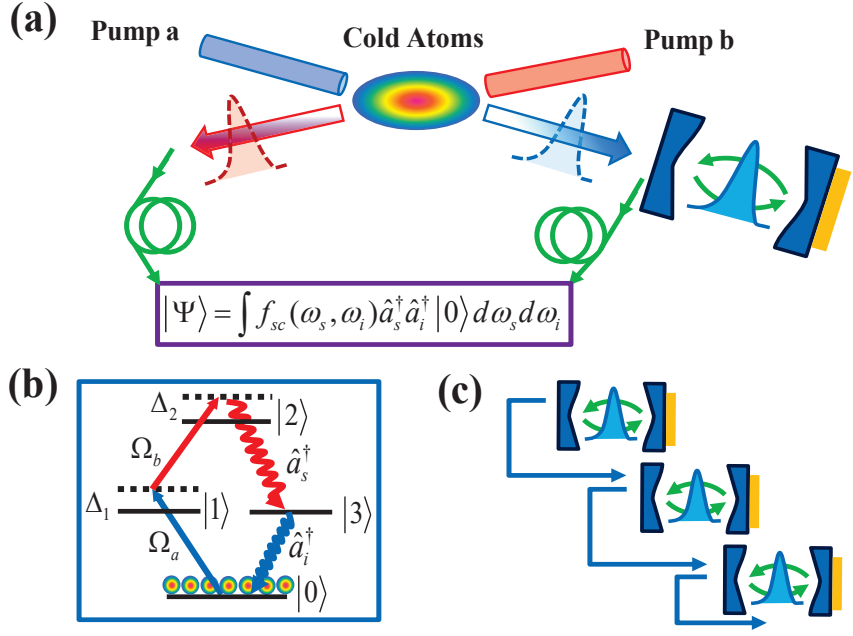}
\caption{Schematic demonstration of spectral compression using an external cavity. (a) The biphoton state ($\hat a_s^\dag$ and $\hat a_i^\dag$) can be generated from a cold atomic ensemble from two counter-propagating pump fields. Its effective spectral function $f_{sc}$ with a spectrally-compressed idler photon can be formed with an external cavity of unit reflectivity at one side. (b) The cascaded atomic levels for biphoton generation under the four-wave mixing condition with two pump ($\Omega_a$ and $\Omega_b$ as Rabi frequencies) and two quantized fields. (c) Further spectral compression of the idler photon under three cavities, as an example.}\label{fig1}
\end{figure}

We consider a biphoton state generated from an atomic ensemble with a number of atoms $N$ through its cascade emissions as shown in Figure \ref{fig1}. The photon pair is highly correlated in their propagating directions with signal and idler wave vectors $\k_{s,i}$ which can be determined by two pump fields $\k_{a,b}$ under the four-wave mixing condition $\delta(\k_a+\k_b-\k_s-\k_i)$. The state vector of the atomic system with a biphoton state under a weak excitation limit becomes
\bea
|\Psi\rangle\approx \ket{0}^{\otimes N}+\frac{\tilde{\Omega}_a \tilde{\Omega}_b g_i^* g_s^*}{4 \Delta_1 \Delta_2} \frac{\sum_{\mu=1}^N e^{i\Delta \textbf{k} \cdot \textbf{r}_\mu}}{\sqrt{2 \pi} \tau} \frac{e^{-(\Delta \omega_s + \Delta \omega_i)^2 \tau^2 / 8}}{\Gamma_3^N/2 - i \Delta \omega_i}|1_{\k_s},1_{\k_i}\rangle,\label{Psi}
\eea
where the pulse areas of Gaussian wave packets are $\tilde\Omega_{a,b}$ with a duration $\tau$, coupling constants $g_{s,i}$, laser detunings $\Delta_{1,2}$, signal and idler's spectral distributions $\Delta\omega_{s,i}$ and the superradiant constant of the idler photon $\Gamma_3^N$ \cite{Chaneliere2006, Jen2012}. The phase-matching condition of four-wave mixing can be seen in the summation of Eq. (\ref{Psi}) in a large $N$ limit, and the derivation of Eq. (\ref{Psi}) is described in detail in Methods section. 
 
The effective biphoton state in frequency spaces reads
\bea
|\Psi_b\rangle=\mathcal{N}\int\int f_b(\omega_{s},\omega_{i})\hat{a}_s^{\dag}(\omega_{s})\hat{a}_i^{\dag}(\omega_{i})|\rm{vac}\rangle d\omega _{s}d\omega_{i},
\eea
where $\mathcal{N}$ and $|\rm{vac}\rangle$ denote a normalization to the spectral density and vacuum photon states, respectively, and the spectral function is defined as  
\bea
f_b(\omega_{s},\omega_{i})\equiv \frac{e^{-(\Delta \omega_s + \Delta \omega_i)^2 \tau^2 / 8}}{\Gamma_3^N/2 - i \Delta \omega_i}.\label{fb}
\eea
The idler photon presents a Lorentzian profile with a full width at half maximum (FWHM) of $\Gamma_3^N$, while the joint Gaussian profile entangles signal and idler photons in frequency spaces \cite{Jen2016a} at an anti-correlation direction. 

The spectrally compressed idler photon to match the absorbing medium can be achieved first by going through a lossless and near-resonant cavity \cite{Seidler2020}, leading to a modified spectral function,     
\bea
f_{\rm m}=f_b(\omega_{s},\omega_{i})\times \left[-\frac{\Gamma_c+i2\Delta\omega_i}{\Gamma_c-i2\Delta\omega_i}\right],\label{fm}
\eea
where $\Gamma_c$ denotes the cavity linewidth, and the above so-called transfer function in brackets \cite{Agarwal1994, Srivathsan2014, Liu2014} can be derived from a two-mirror cavity with one side of unit reflectivity. Under this lossless transformation, we next apply a phase compensation to the temporal wave packet to essentially remove the time-dependent phases, and we obtain the compressed spectral function as 
\bea
f_{\rm sc} =  F\left[\left|F^{-1}\left[f_{\rm m}(\omega_{s},\omega_{i})\right]\right|\right],\label{fsc}
\eea 
where $F$ represents the Fourier transform, and taking the absolute value above means a total removal of the phases in time. The temporal wave packet can be derived as 
\bea
F^{-1}\left[f_{\rm m}(\omega_{s},\omega_{i})\right]=\frac{2\sqrt{2\pi}}{\tau}e^{-2t_s^2/\tau^2}\frac{(\Gamma_3^N+\Gamma_c)e^{-\Gamma_3^N(t_i-t_s)/2}-2\Gamma_c e^{-\Gamma_c(t_i-t_s)/2}}{\Gamma_3^N-\Gamma_c}\Theta(t_i-t_s),\label{temporal}
\eea
where $\Theta$ is the Heaviside function, presenting a causal relation that the idler photon emits after the signal. The abrupt phase change in Eq. \ref{temporal} takes place at $t_i-t_s$ $=$ $2\log[(\Gamma_3^N+\Gamma_c)/(2\Gamma_c)]/(\Gamma_3^N-\Gamma_c)$ which has been obtained in single photon spectral compression \cite{Seidler2020}. It is this removal of abrupt phase change in time as shown in Eq. (\ref{fsc}) that enables the spectral compression. Here we present a generalized biphoton state under a spectral compression, from which we are able to further obtain the useful information of continuous entanglement entropy along with the compressed FWHM of the idler photon.   


\subsection*{Continuous entanglement entropy and FWHM of idler photon}

From Eq. (\ref{fsc}), we numerically calculate the entanglement entropy $S$ in frequency spaces using Schmidt decomposition \cite{Law2000, Parker2000}. The theoretical background of Schmidt decomposition is reviewed and presented in Methods section, which enables the quantification of an entangled biphoton source. As long as $S>0$, it indicates an entangled biphoton state, where $S$ $\equiv$ $-\sum_{n=1}^{\infty}\lambda_{n}\log_2\lambda_{n}$ with eigenvalues $\lambda_n$ of the $n$th Schmidt mode for a specific spectral function. $S$ vanishes and corresponds to a separable state only when $\lambda_1=1$ under the normalization of the state probability $\sum_n\lambda_n=1$. 

\begin{figure}[t]
\centering
\includegraphics[width=12cm,height=6cm]{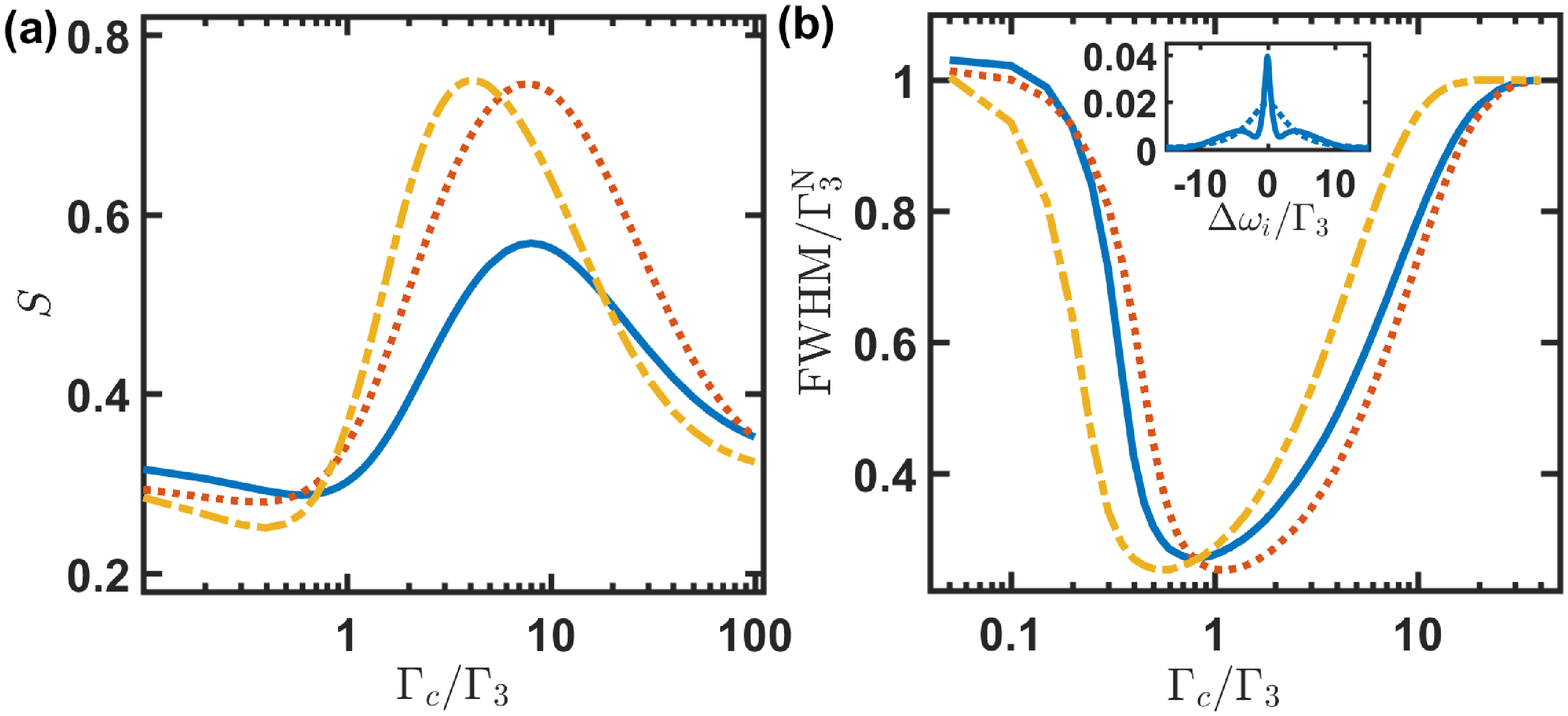}
\caption{Continuous entanglement entropy $S$ and FWHM of the idler photon. Under the spectrally compressed biphoton state with a spectral function $f_{\rm{sc}}$, the entanglement entropy $S$ in frequency spaces and FWHM of the idler photon at $\Delta\omega_s=0$ are plotted respectively in (a) and (b), for ($\Gamma_3\tau$, $\Gamma_3^N/\Gamma_3$) $=$ ($0.25$, $5$) (solid-blue), ($0.25$, $10$) (dotted-red), and ($0.5$, $5$) (dash-dotted yellow) in both plots. $S$ is numerically obtained by taking the range of $\pm 300\Gamma_3$ in the distributions of $\Delta\omega_{s,i}$. The inset in (b) presents the case of the idler spectral distribution for ($\Gamma_3\tau$, $\Gamma_3^N/\Gamma_3$) $=$ ($0.25$, $5$) at $\Gamma_c/\Gamma_3$ $=$ $0.8$ (solid-blue) and $15$ (dashed-blue).}\label{fig2}
\end{figure}

In Figure \ref{fig2}(a), we plot the entanglement entropy $S$ for $f_{\rm sc}$ in Eq. (\ref{fsc}) as we vary the cavity linewidth $\Gamma_c$. The relatively low values of $S$ are associated with spectrally compressed idler photon shown in Figure \ref{fig2}(b), where most prominently compressed idler photon occurs at $\Gamma_c$ $\lesssim$ $\Gamma_3$ with a reduction of $\sim$ $75\%$ of $\Gamma_3^N$. As a comparison to the spectral function $f_{b}$ without a cavity in Eq. (\ref{fb}), the corresponding $S$ are $1.33$, $2.44$ and $2.04$ \cite{Jen2012-2} respectively for the parameters ($\Gamma_3\tau$, $\Gamma_3^N/\Gamma_3$) considered in Figure \ref{fig2}. As expected, both $S$ and FWHM at $\Gamma_c\rightarrow\infty$ in Figure \ref{fig2} approach the cases at $\Gamma_c\rightarrow 0$, since the transfer function in Eq. (\ref{fm}) in respective limits makes an equivalent effect on the idler photons. In determining the FWHM in Figure \ref{fig2}(b), we take the window of $\Delta\omega_i$ that contains $50\%$ of the total spectral energy as a measure \cite{Seidler2020}, since the idler wave packet emerges with multiple peaks at small $\Gamma_c$ in particular, in contrast to the case with a smooth and Lorentzian profile at a relatively large $\Gamma_c$. This can be seen in the inset of Figure \ref{fig2}(b), as an example.  

The minimum of $S$ in Figure \ref{fig2}(a) approximately corresponds to the minimum of FWHM. This can be explained by $f_b$ in Eq. (\ref{fb}), where an effectively smaller $\Gamma_3^N$ of idler photons limits the distribution of the joint Gaussian wave packets at an anti-correlation direction, leading to a reduction of frequency entanglement. A maximal $S$ emerges instead, which can be attributed to a relatively weak effect from phase compensation. We note that the values at very small $\Gamma_c$ in Figure \ref{fig2}(b) do not converge completely to $\Gamma_3^N$ owing to the extremely large scale of $\Gamma_c^{-1}$ in time that could not be faithfully covered in the numerical calculations of inverse Fourier transforms. This long-time behavior can be seen in Eq. (\ref{temporal}) when $\Gamma_c\rightarrow 0$.   


\subsection*{Multiple cavities}

Next, we consider a scheme which employs multiple cavities as shown in Figure \ref{fig1}(c) to further enhance the effect of spectral compression on the idler photons. The effective spectral distribution for idler photons passing through multiple cavities becomes 
\bea
f_{\rm{sc}}^n=F\left[\left|F^{-1}\left[f_{\rm{sc}}^{n-1}\times \left(-\frac{\Gamma_c+i2\Delta\omega_i}{\Gamma_c-i2\Delta\omega_i}\right)\right]\right|\right],
\eea
where $n\geq 2$ represents the number of cavities, and $f_{\rm sc}^1\equiv f_{\rm sc}(\tau=0)$. Under the condition of $\tau=0$ in $f_{\rm sc}$, we consider only the idler's spectral compression without the influence of the signal, which is analogous to an implementation of heralded single photon conditioning on the detection of the signal. 

\begin{figure}[t]
\centering
\includegraphics[width=12cm,height=6cm]{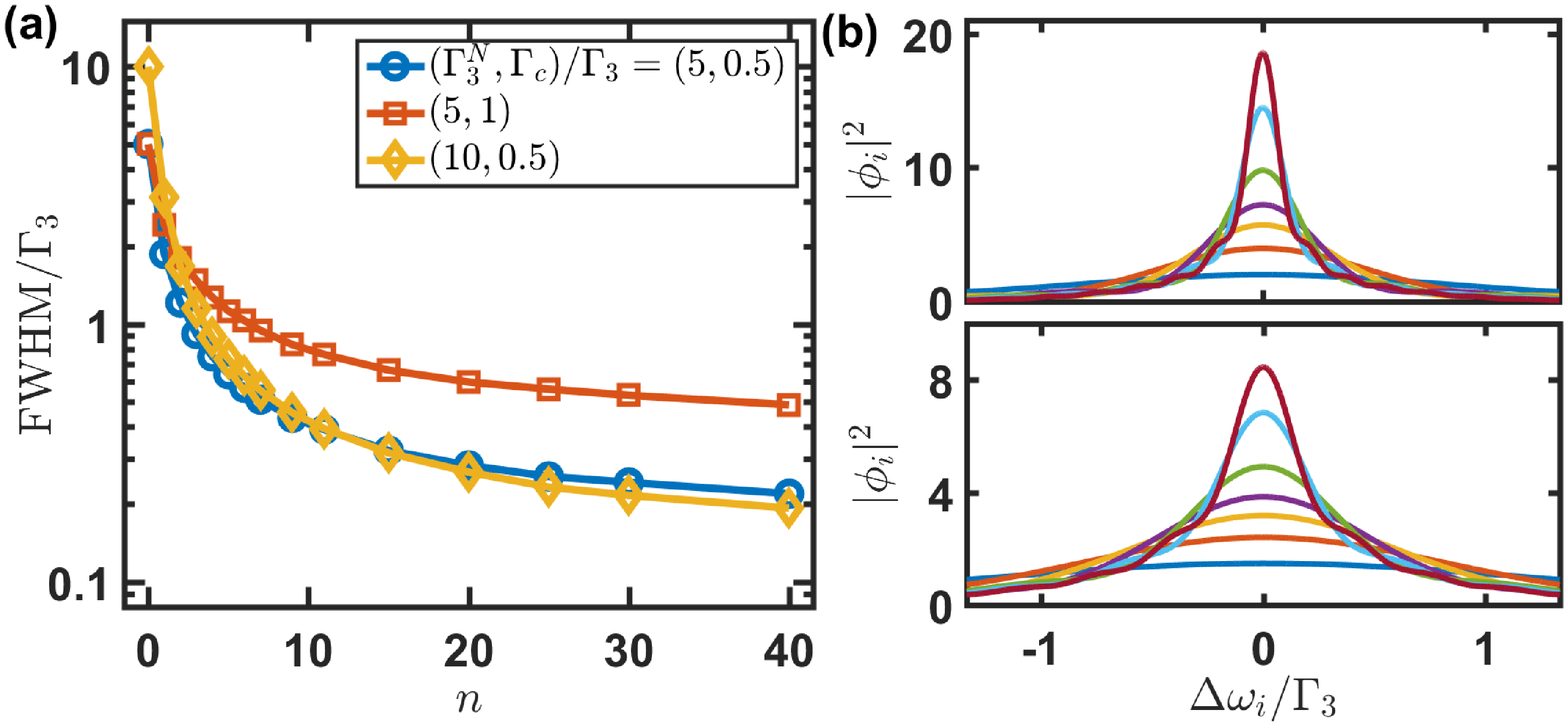}
\caption{FWHM of heralded idler photons under multiple cavities. (a) FWHM of the idler photon wave packets passing through $n$ lossless and near-resonant cavities. (b) The corresponding wave packets $|\phi_i|^2$ are shown in the upper and lower plots for $(\Gamma_3^N,\Gamma_c)/\Gamma_3$ $=$ $(5,0.5)$ and $(5,1)$, respectively. The idler profiles become more compressed spectrally as $n$ increases from $1$, $3$, $5$, $7$, $11$, $20$ to $30$.}\label{fig3}
\end{figure}

In Figure \ref{fig3}, we show the linewidth of heralded idler photons after passing through the cavities. This compressed FWHM develops from the time-dependent phase removals every time before the phase modulation from the next cavity in frequency space, which provides an effectively lower bound of FWHM that this scheme can achieve. The linewidth decreases significantly as the number of cavities $n\lesssim 10$ and saturates as more $n$ is used. For $n\sim 7$, the compressed idler photon wave packets can reach better than $10\%$ of the original linewidth without cavities near the chosen optimal $\Gamma_c$, demonstrating the improved capability of spectral compression with multiple cavities.      

\begin{figure}[t]
\centering
\includegraphics[width=12cm,height=6cm]{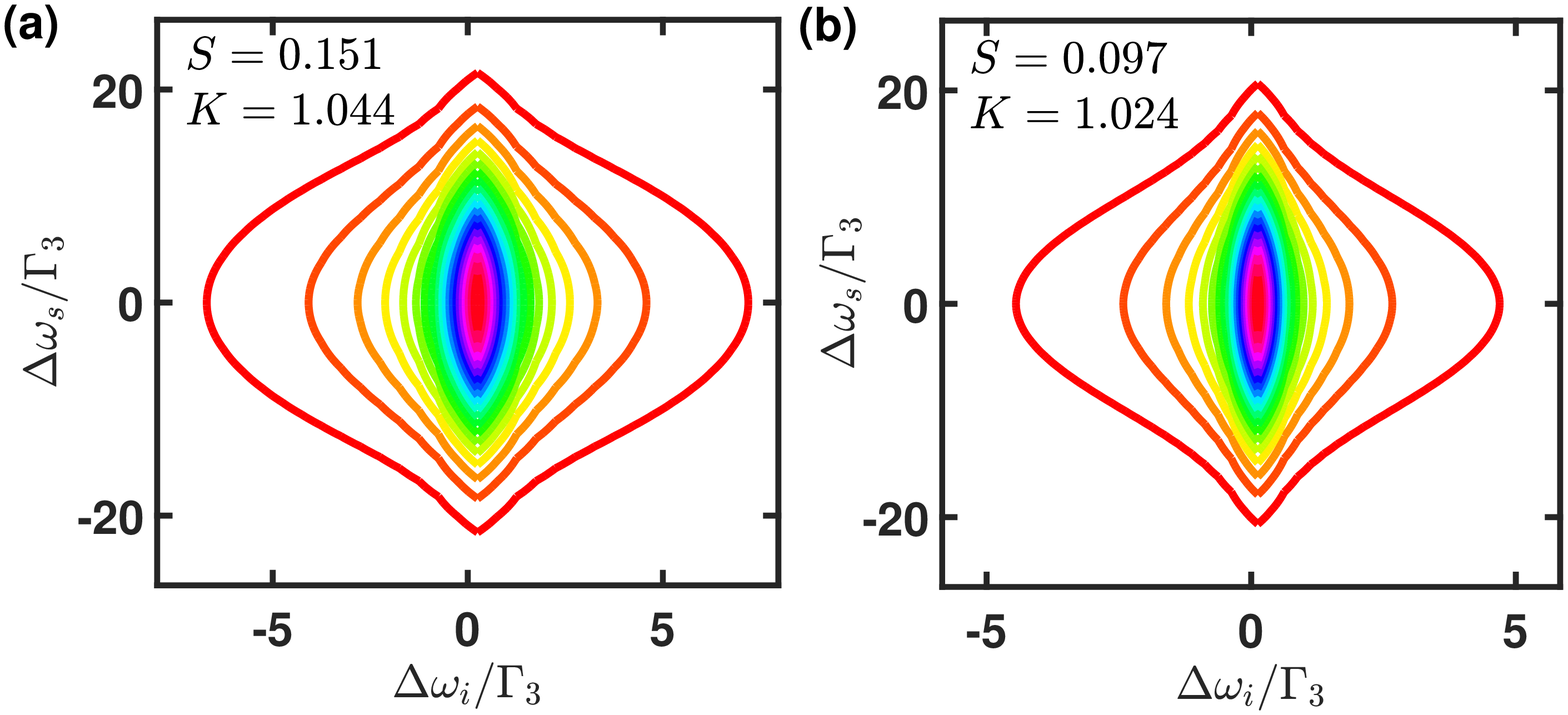}
\caption{Spectral distributions from a compressed idler photon after $n=7$ cavities. Effective $f_{\rm sc}$ is plotted for $\Gamma_3\tau=0.25$ and (a) $\Gamma_3^N/\Gamma_3=0.95$ and (b) $0.51$. We note that $\Gamma_3^N$ here is not superradiant anymore after passing through multiple cavities and serves as an effective compressed linewidth. These compressed FWHMs are extracted respectively from the cases of $(\Gamma_3^N,\Gamma_c)/\Gamma_3$ $=$ $(5,0.5)$ and $(5,1)$ for $n=7$ in Figure \ref{fig3}. We numerically calculate the values of $S$ and $K$ using the same spectral ranges in Figure \ref{fig2}. The estimated relative errors to the projected values $a$ from grid sizes $x$ of $\Delta\omega_{s,i}$ in obtaining $S$ and $K$ are $\lesssim 9\%$ and $0.5\%$, respectively, by fitting $S$ and $K$ with $a+b*x$ with arbitrary parameters $a$ and $b$.}\label{fig4}
\end{figure}

Finally, in Figure \ref{fig4}, we show the biphoton spectral distributions for the case of $n=7$ using the effective FWHM obtained in Figure \ref{fig3}. This presents again the lower bound of entanglement entropy $S$ the biphoton state can possess. We also calculate the associated Schmidt number $K=1/\sum_n\lambda_n^2$ \cite{Grobe1994, Law2004} which specifies an average number of correlated Schmidt modes in the biphoton source. These small values of $S$ and $K$ emerge from spectrally compressed idler photon wave packets, which manifests an approximate pure single photon source, applicable in implementing quantum networks using linear optical setups.   

\section*{Discussion}

To efficiently generate correlated biphoton states in the cascade atomic configurations, a large and dense atomic medium is required to fulfill a phase matching between the emitted signal, idler photons and two external weak pumps under the four-wave mixing condition. In this optically-dense medium, a superradiant idler photon emerges as an enhanced radiation phenomenon owing to the collective resonant dipole-dipole interactions. This broadened linewidth of lower infrared transition, however, does not match the atomic intrinsic transition, leading to an inefficient light qubit transfer to the absorbing quantum interface using neutral atoms. The spectral compression presented here using multiple cavities can actively manipulate the frequency entanglement properties and essentially operate a spectral shaping in the cascaded biphoton source. Significant spectral compression more than one tenth of the original linewidth can be reached using only a couple of lossless and near-resonant cavities under an optimal cavity linewidth. The scheme of spectral compression with cavities demonstrates a flexible control over the photon linewidth, which can potentially facilitate the construction of hybrid quantum systems \cite{Kuriski2015, Clerk2020} to take full account of their complimentary functionalities. In addition, the heralded pure single photon sources can be generated by reducing the frequency entanglement of the spectrally compressed biphoton, which are essential in the development of photon-photon quantum logic gates \cite{Li2021} in an optical quantum computer.      

\section*{Methods}

\subsection*{Theoretical model}

As shown in Figure \ref{fig1}, two pump fields of Rabi frequencies $\Omega_{a,b}$ drive the system from the ground state $|0\rangle\rightarrow |1 \rangle$ and $|1 \rangle\rightarrow |2\rangle$, respectively. Under the four-wave mixing condition, highly correlated photons $\hat a_s$ and $\hat a_i$ are spontaneously emitted via $|3\rangle$ and back to $|0\rangle$. The Hamiltonian in interaction picture reads \cite{Jen2016a}
	\bea
	\centering
	V_I &=& - \sum_{m=1,2} \Delta_m \sum_{\mu=1}^{N} \ket{m}_{\mu} \bra{m} - \sum_{m=a,b} \left(\frac{\Omega_m}{2} \hat{P}_m^\dagger + h.c.\right) \nonumber\\
	&-& i\sum_{m=s,i} \left[ \sum_{\textbf{k}_m,\lambda_m} g_m \hat{a}_{\textbf{k}_m,\lambda_m} \hat{Q}_m^\dagger e^{-i\Delta \omega_m t} - h.c. \right], 
	\eea
where we let $\hslash$ $=$ $1$, the laser detunings are denoted as $\Delta_1 = \omega_a - \omega_1$ and $\Delta_2 = \omega_a + \omega_b - \omega_2$ with the atomic transition energies $\omega_{m = 1, 2, 3}$ and the central frequencies of pump and emitted signal and idler fields $\omega_{a,b,s,i}$, and the polarization and wave vectors are $\lambda_m$ and $\textbf{k}_m$, respectively. The signal and idler coupling constants $g_m$ have absorbed $\epsilon_{\textbf{k}_m, \lambda_m} \cdot \hat{d}_m^*$ with the polarization direction $\epsilon_{\textbf{k}_m, \lambda_m}$ for quantized bosonic fields $\hat{a}_{\textbf{k}_m,\lambda_m}$ and unit direction of dipole operators $\hat{d}_m$. Various dipole operators are defined as $\hat{P}_a^\dagger \equiv \sum_\mu \ket{1}_\mu \bra{0} e^{i \textbf{k}_a \cdot \textbf{r}_\mu }$, $\hat{P}_b^\dagger \equiv \sum_\mu \ket{2}_\mu \bra{1} e^{i \textbf{k}_b \cdot \textbf{r}_\mu }$, $\hat{Q}_s^\dagger \equiv \sum_\mu \ket{2}_\mu \bra{3} e^{i \textbf{k}_s \cdot \textbf{r}_\mu }$, $\hat{Q}_i^\dagger \equiv \sum_\mu \ket{3}_\mu \bra{0} e^{i \textbf{k}_i \cdot \textbf{r}_\mu }$. 

Under the limit of weak excitation fields that $\sqrt{N} \abs{\Omega_a} \ll \Delta_1$, a single excitation Hilbert space is sufficient. The state function of the system becomes   
	\bea
		\ket{\psi(t)} &=& \mathcal{E}(t) \ket{0, \rm{vac}} + \sum_{\mu = 1}^{N} A_\mu(t) \ket{1_\mu, \rm{vac}} + \sum_{\mu = 1}^{N} B_\mu(t) \ket{2_\mu, \rm{vac}} \nonumber\\
		&+& \sum_{\mu = 1}^{N} \sum_{s} C_s^\mu(t) \ket{3_\mu, 1_{\textbf{k}_s, \lambda_s}} + \sum_{s,i} D_{s,i}(t) \ket{0, 1_{\textbf{k}_s, \lambda_s}, 1_{\textbf{k}_i, \lambda_i}}, 
	\eea
where the collective single excitation states and the vacuum photon state are $\ket{m_\mu} \equiv \ket{m_\mu} \ket{0}_{\nu \neq \mu}^{\otimes N-1}$ and $\ket{\rm{vac}}$, respectively. We then apply Schr\"{o}dinger equation $i\hslash \frac{\partial}{\partial t} \ket{\psi(t)} = V_I (t) \ket{\psi(t)}$, and the coupled equations of motion can be obtained,  
	\bea
		 i \dot{\mathcal{E}} &=& - \frac{\Omega^\ast_a}{2} \sum_\mu e^{-i \textbf{k}_a \cdot \textbf{r}_\mu} A_\mu, \\
		 i \dot{A}_\mu &=& - \frac{\Omega_a}{2} e^{i \textbf{k}_a \cdot \textbf{r}_\mu} \mathcal{E} - \frac{\Omega^\ast_b}{2} e^{-i \textbf{k}_b \cdot \textbf{r}_\mu} B_\mu - \Delta_1 A_\mu, \\
		 i \dot{B}_\mu &=& - \frac{\Omega_b}{2} e^{-i \textbf{k}_b \cdot \textbf{r}_\mu} A_\mu - \Delta_2 B_\mu - i \sum_{k_s, \lambda_s} g_s e^{i \textbf{k}_s \cdot \textbf{r}_\mu} e^{-i (\omega_s - \omega_{23} - \Delta_2) t} C^\mu_s, \\
		 \dot{C^\mu_{s,i}} &=& i g^\ast_s e^{-i \textbf{k}_s \cdot \textbf{r}_\mu} e^{i (\omega_s - \omega_{23} - \Delta_2) t} B_\mu - i \sum_{k_i, \lambda_i} g_i e^{i \textbf{k}_i \cdot \textbf{r}_\mu} e^{-i (\omega_i - \omega_3) t} D_{s,i}, \label{C}\\
		 i \dot{D_{s,i}} &=& i g^\ast_i \sum_\mu e^{-i \textbf{k}_i \cdot \textbf{r}_\mu} e^{i (\omega_i - \omega_3) t} C^\mu_s,\label{D}
	\eea
where we can safely ignore the spontaneous decays during the pumping process owing to large detunings $\Delta_{1,2}$. Under the adiabatic approximation that the system follows the excitation fields, we are able to obtain the leading order solutions of $\mathcal{E} \thickapprox 1$, $A_\mu(t)$ $\approx$ $-\Omega_a(t)e^{i\k_a\cdot\r_\mu}/(2\Delta_1)$ and $B_\mu(t)$ $\approx$ $\Omega_a(t)\Omega_b(t)e^{i(\k_a+\k_b)\cdot\r_\mu}/(4\Delta_1\Delta_2)$ from the steady-state solutions of $\dot{\mathcal{E}} = \dot{A}_\mu = \dot{B}_\mu = 0$. 

Substituting the above solutions into Eqs. (\ref{C}) and (\ref{D}), and considering a symmetrical excitation state, $(\sqrt{N})^{-1}$ $\sum_{\mu=1}^N$ $e^{i(\k_a+\k_b-\k_s)\cdot\r_\mu}|3\rangle_\mu|0\rangle^{\otimes N-1}$, we obtain the biphoton state $|1_{\k_s},1_{\k_i}\rangle$ in a large $N$ limit,
\bea
D_{s,i}(t)= g_{i}^{\ast}g_{s}^{\ast}\sum_{\mu=1}^Ne^{i\Delta\k\cdot\r_{\mu}}\int_{-\infty}^{t}\int_{-\infty}^{t^{\prime}}dt^{\prime\prime}dt^{\prime}
e^{i\Delta\omega_{i}t^{\prime}} e^{i\Delta\omega_{s}t^{\prime\prime}} \frac{\Omega_{a}(t'')\Omega_{b}(t'')}{4\Delta_{1}\Delta_{2}}e^{\left(-\Gamma_{3}^N/2+i\delta\omega_{i}\right)(t^{\prime}-t^{\prime\prime})},\label{Dsi2}
\eea
where $\Delta \omega_s \equiv \omega_s - (\omega_2 - \omega_3 +\Delta_2)$, $\Delta \omega_i \equiv \omega_i - \omega_3$. $\sum_{\mu=1}^N e^{i\Delta\k\cdot\r_{\mu}}$ defines the four-wave mixing condition and indicates a phase-matched and highly correlated biphoton state when  $\Delta\k$ $\equiv$ $\k_{a}$ $+$ $\k_{b}$ $-$ $\k_{s}$ $-$ $\k_{i}$ $\rightarrow$ $0$. We have denoted the superradiant decay constant of the idler photon \cite{Chaneliere2006, Jen2012} as $\Gamma_{3}^{\rm N}$ $=$ $(N\bar{\mu}+1)\Gamma_{3}$ with an intrinsic decay constant $\Gamma_3$ and a geometrical constant $\bar{\mu}$ \cite{Rehler1971}. $\Gamma_{3}^{\rm N}$ depends on the density and geometry of the atomic ensemble, and the relevant collective frequency shift $\delta\omega_i$ \cite{Scully2009, Jen2015} from the resonant dipole-dipole interactions \cite{Lehmberg1970} can be renormalized and absorbed into idler's central frequency $\omega_3$. 

We next assume the excitations with Gaussian wave packets $\Omega_{a,b}(t)$ $=$ $\tilde{\Omega}_{a,b}e^{-t^{2}/\tau^{2}}/(\sqrt{\pi}\tau)$ with a pulse duration $\tau$, and in a long time limit, we finally obtain the probability amplitude of the biphoton state,
	\bea
	D_{si} (\Delta \omega_s, \Delta \omega_i) = \frac{\tilde{\Omega}_a \tilde{\Omega}_b g_i^* g_s^*}{4 \Delta_1 \Delta_2} \frac{\sum_{\mu=1}^N e^{i\Delta \textbf{k} \cdot \textbf{r}_\mu}}{\sqrt{2 \pi} \tau} f_b(\omega_s, \omega_i), ~ f_b(\omega_s, \omega_i) \equiv \frac{e^{-(\Delta \omega_s + \Delta \omega_i)^2 \tau^2 / 8}}{\Gamma_3^N/2 - i \Delta \omega_i}. \label{eq: fc}
	\eea
The Gaussian distribution in $f_b(\omega_s, \omega_i)$ prefers $\Delta \omega_s + \Delta \omega_i = 0$ which shows an anti-correlation between the signal and idler photons under the energy conservation of pumps and biphoton emissions $\omega_s + \omega_i = \omega_a + \omega_b$.

\subsection*{Schmidt decomposition}

To calculate the entanglement entropy in continuous frequency spaces \cite{Branning1999, Law2000, Parker2000, Jen2012-2}, here we introduce Schmidt decomposition analysis useful for bipartite systems. For a spectrally entangled source of signal $\hat a_{\lambda_{s}}$ and idler $\hat a_{\lambda_{i}}$ photons with some polarizations $\lambda_{s}$ and $\lambda_{i}$, respectively, the biphoton state $|\Psi'\rangle$ with a spectral function $f_b'(\omega_{s},\omega_{i})$ reads
\bea
|\Psi'\rangle=\int f_b'(\omega_{s},\omega_{i})\hat{a}_{\lambda_{s}}^{\dag}(\omega_{s})\hat{a}_{\lambda_{i}}^{\dag}(\omega_{i})|0\rangle d\omega
_{s}d\omega_{i}.
\eea
We can calculate the entanglement entropy in the Schmidt bases, where the state vectors can be written as
\bea
|\Psi'\rangle=\sum_{n}\sqrt{\lambda_{n}}\hat{b}_{n}^{\dag}\hat{c}_{n}^{\dag}|0\rangle,~
\hat{b}_{n}^{\dag}\equiv\int\psi_{n}(\omega_{s})\hat{a}_{\lambda_{s}}^{\dag}(\omega_{s})d\omega_{s},~
\hat{c}_{n}^{\dag}\equiv\int\phi_{n}(\omega_{i})\hat{a}_{\lambda_{i}}^{\dag}(\omega_{i})d\omega_{i}.
\eea
The effective creation operators $\hat{b}_{n}^{\dag}$ and $\hat{c}_{n}^{\dag}$ associate with the eigenmodes $\psi_{n}$ and $\phi_{n}$ respectively, and $\lambda_n$'s are the eigenvalues and probabilities for the $n$th eigenmode. These eigenmodes can be obtained by
\begin{eqnarray}
&&\int K_{1}(\omega,\omega^{\prime})\psi_{n}(\omega^{\prime})d\omega^{\prime}  =\lambda_{n}\psi_{n}(\omega),\\
&&\int K_{2}(\omega,\omega^{\prime})\phi_{n}(\omega^{\prime})d\omega^{\prime}  =\lambda_{n}\phi_{n}(\omega),
\end{eqnarray}
where the kernels for one-photon spectral correlations \cite{Law2000, Parker2000} can be constructed as
\begin{eqnarray}
&&K_{1}(\omega,\omega^{\prime}) \equiv\int f_b'(\omega,\omega_{1})f_b'^{\ast}(\omega^{\prime},\omega_{1})d\omega_{1},\\
&&K_{2}(\omega,\omega^{\prime}) \equiv\int f_b'(\omega_{2},\omega)f_b'^{\ast}(\omega_{2},\omega^{\prime})d\omega_{2}. 
\end{eqnarray}
The orthogonality of these eigenmodes can be satisfied as $\int\psi_{i}(\omega)$$\psi_{j}^*(\omega)d\omega$ $=$ $\delta_{ij}$ and $\int\phi_{i}(\omega)$$\phi_{j}^*(\omega)d\omega$ $=$ $\delta_{ij}$, and the normalization of Schmidt analysis requires $\sum_{n}\lambda_{n}$ $=$ $1$.

We can calculate the entanglement entropy $S$ in the Schmidt bases as
\begin{equation}
S=-\sum_{n=1}^{\infty}\lambda_{n}\textrm{log}_2\lambda_{n},\label{entropy}
\end{equation}
where a separable (non-entangled) state with $\lambda_{1}=1$ gives a null $S$, and a finite bipartite entanglement of $S>0$ corresponds to more than one Schmidt numbers $\lambda_n$ in $|\Psi'\rangle$. Another associated value of Schmidt number $K=1/\sum_n \lambda_{n}^2$ \cite{Grobe1994, Law2004} is useful to provide an averaging measure of biphoton correlations in continuous frequency spaces. $K>1$ means an entangled biphoton source, and $K$ itself represents the number of correlated Schmidt modes, which is maximal when the volume of accessible modes are equally accessed. 

\section*{Data availability}

Data available on reasonable request from the corresponding author.


\section*{Acknowledgments}
We acknowledge the support from the Ministry of Science and Technology (MOST), Taiwan, under the Grant No. MOST-109-2112-M-001-035-MY3.

\section*{Author contributions}

H.H.J. conceived the idea, organized the collaborations, and wrote the manuscript with contributions from the other authors; Y.E.W. and N.Y.T. conducted the numerical calculations, W.S.H. derived the equations, and all authors analyzed the results.

\section*{Competing interest}

The authors declare that they have no competing interests. 

\end{document}